\documentclass[aip,twocolumn,showpacs,preprintnumbers,showkeys]{revtex4}
\usepackage{epsfig}
\usepackage{color}
\usepackage{amsmath}
\usepackage{amsfonts}
\usepackage{amssymb}
\usepackage{graphicx}
\allowdisplaybreaks[4]
\usepackage{natbib}
\def\be{\begin{equation}}
\def\ee{\end{equation}}
\def\bdm{\begin{displaymath}}
\def\edm{\end{displaymath}}

\begin{document}

\title{Constraints for the aperiodic O-mode streaming instability}
\author{M. Lazar$^{1,2}$,  R. Schlickeiser$^{2,3}$, S. Poedts$^{1}$, A. Stockem$^{2}$ and S. Vafin$^2$}
\email{mlazar@tp4.rub.de} \affiliation{$^1$ Center for Plasma
Astrophysics, Celestijnenlaan 200B, 3001 Leuven, Belgium \\
$^2$ Institut f\"ur Theoretische Physik, Lehrstuhl IV: Weltraum- und
Astrophysik, Ruhr-Universit\"at Bochum, D-44780 Bochum, Germany\\
$^3$ Research Department Plasmas with Complex Interactions,
Ruhr-Universit\"at Bochum, D-44780 Bochum, Germany }
\date{\today}

\begin{abstract}

In plasmas where the thermal energy density exceeds the magnetic
energy density ($\beta_\parallel > 1$), the aperiodic ordinary mode
(O-mode) instability is driven by an excess of parallel temperature
$A = T_\perp /T_\parallel < 1$ (where $\parallel$ and $\perp$ denote
directions relative to the uniform magnetic field). When stimulated
by parallel plasma streams the instability conditions extend to low
beta states, i.e., $\beta_\parallel <1$, and recent studies have
proven the existence of a new regime, where the anisotropy threshold
decreases steeply with lowering $\beta_\parallel \to 0$ if the
streaming velocity is sufficiently high. However, the occurrence of
this instability is questionable especially in the low-beta plasmas,
where the electrostatic two-stream instabilities are expected to
develop much faster in the process of relaxation of the
counterstreams. It is therefore proposed here to identify the
instability conditions for the O-mode below those required for the
onset of the electrostatic instability. An hierarchy of these two
instabilities is established for both the low $\beta_\parallel <1$
and large $\beta_\parallel > 1$ plasmas. The conditions where the
O-mode instability can operate efficiently are markedly constrained
by the electrostatic instabilities especially in the low-beta
plasmas.

\end{abstract}


\maketitle

\section{Introduction}

There is an increased interest for understanding the mechanisms that
can destabilize the aperiodic or weakly propagating modes in
anisotropic plasmas. Of these the purely growing (aperiodic, i.e.,
$\Re(\omega) = 0$) ordinary (O) mode instability has recently
received particular attention owing to its potential applications in
space plasmas \cite{st06, ib13, ha14, sc14}. In a plasma at rest the
O-mode instability (OMI) can develop only if the plasma beta is
sufficiently high ($\beta \equiv 8 \pi n k_B T /B_0^2 > 2$), and
thermal energy is higher in the direction parallel to the uniform
magnetic field (${\bf B}_0$), i.e., $A = T_{\perp}/T_{\parallel} <
1$ \cite{ha68, da70, ib12,la13}.

In the presence of streams, propagating along the ordered magnetic
field, the activity of this instability extends to low beta $\beta <
1$ states \cite{ib13, ha14, sc14, bo70, le71, ga72} (although the
instability is inhibited by the magnetic field by limiting the range
of unstable wavenumbers \cite{st06}). Thus, for two symmetric
counterstreams of electrons (subscript $e$) the conditions necessary
for the O-mode instability are \cite{bo70} $\beta_{e,\parallel}>2/
(1+2V_e^2 /u_{e,\parallel}^2)$ and $A_e < 1+2V_e^2/
u_{e,\parallel}^2$, where $u_{e,\parallel}=(2k_B T_{e,\parallel}
/m_e)^{1/2}$ is the electron thermal velocity in parallel direction,
$\beta_{e,\parallel} = 8 \pi n k_B T_{e,\parallel} /B_0^2$, and
$V_e$ is the streaming velocity. (Symmetric counterstreams enable to
analyze the O-mode decoupled from the extraordinary (X) mode, which
is less susceptible to the instability. Fig.~\ref{f1} presents a
schematic with the possible configurations of symmetric
counterstreams.) These conditions simply show that instability is
also predicted for low values of $\beta_{e,\parallel} <1$ (if
streaming velocity is large enough $V_e
> u_{e,\parallel}/\sqrt{2}$), and for high values of the
anisotropy $A_e > 1$ (given by an excess of perpendicular
temperature). Growth rates are of the order of electron cyclotron
frequency $|\Omega_e|$.

A plasma system with counterstreaming ions can be more susceptible
to the O-mode instability than one in which only electrons are
streaming \cite{le71,ga72}. The streaming ions enlarge the range of
unstable wavenumbers but affect only slightly the maximum growth
rates. However, for large growth rates (of the order of
$|\Omega_e|$), high electron beta $\beta_{e,\parallel} \sim 1$ and
temperature anisotropy in parallel direction, i.e., $T_{e,\parallel}
\gg T_{e,\perp}$, are needed. On the other hand, only sufficiently
high streaming velocities $V_e > u_{e,\parallel}$ can predict the
occurrence of instability at large $A_e > 1$.

Recently, a number of studies have been devoted to express
analytically the marginal condition of the O-mode instability
\cite{ib12,ib13,ha14, sc14}. It is now straightforward to determine
the instability conditions for the whole range of plasma beta,
including the low-beta regime where the O-mode is driven unstable
only by the relative motion of the plasma streams \cite{ib13,sc14}. 
For low $\beta_{\parallel} < 1$ it is shown for the first
time the existence of a new regime, where the anisotropy thresholds
steeply decrease with lowering $\beta_\parallel \to 0$ if the
streaming velocity is sufficiently high. The exact instability
thresholds have been derived numerically for small but finite growth
rates \cite{ha14}, in order to confirm the marginal instability
condition in analytical forms, and implicitly the existence of the
new regime in the low-beta limit.

Counterstreaming plasmas are also subject to the electrostatic
two-stream instabilities (TSI), e.g., electron-electron,
electron-ion and ion-ion, of which, the instabilities driven by
electrons (with a growth rate of the order of electron plasma
frequency $\omega_{pe} = \sqrt{4\pi n e^2/m_e}$) appear to be faster
\cite{st64,la02}. Moreover, the electrostatic two-stream instability
is in most scenarios faster than the O-mode instability (OMI),
except for streaming velocities very near or below the threshold for
the onset of the two-stream instability \cite{le71, ga72}. The
instability of the O-mode remains to be established only for
streaming velocities below the threshold of the two-stream
instability even for low $\beta_{e,\parallel} < 1$. In this paper we
propose to delimitate these regimes on the basis of the results in
Ref.~13, where the marginal condition has been derived
systematically for different types and characteristics of two-stream
instabilities. Thus, in Sec.~II we revisit the instability
conditions of the O-mode for three specific cases of symmetric
counterstreams. These are confronted in Sec.~III with the
instability conditions of the electrostatic two-stream modes,
providing the existence conditions for the O-mode streaming
instability. The new criteria are discussed along with our final
conclusions in the last section.

\section{The O-mode instability}

\begin{figure}
\centerline{\includegraphics[width=85mm]{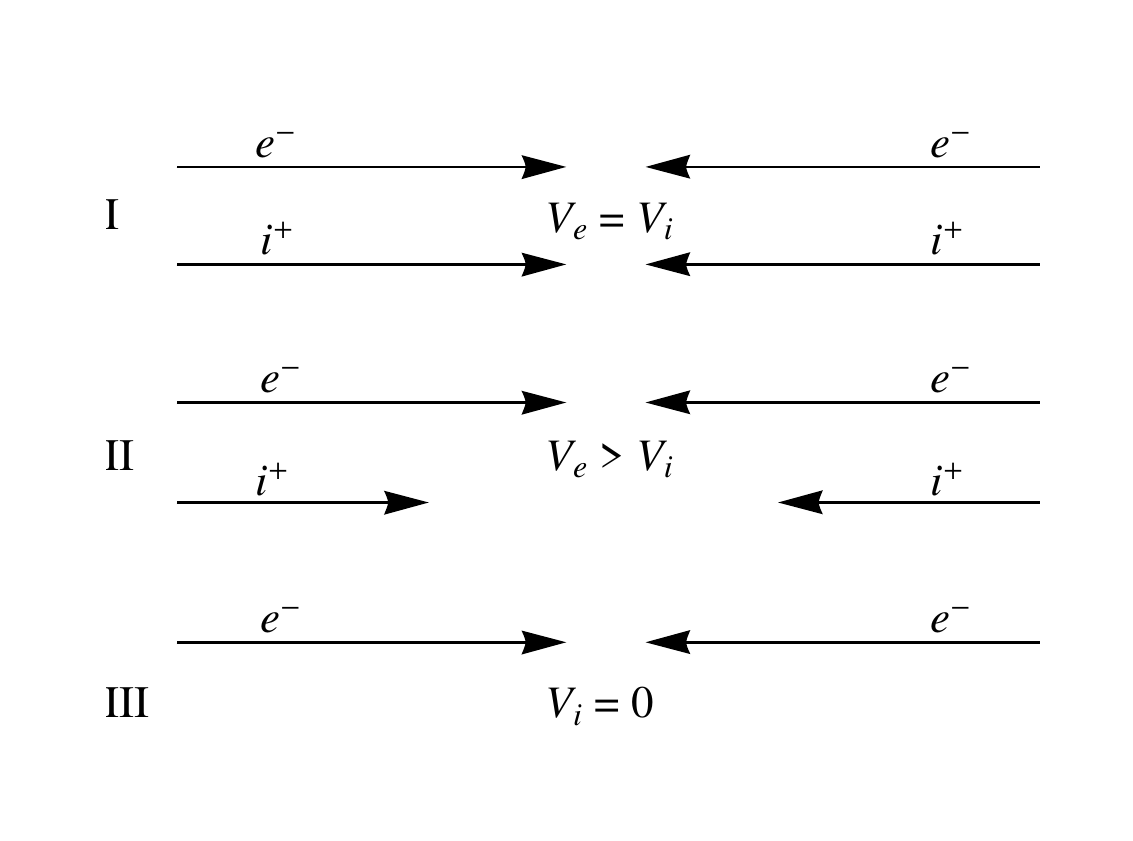}}
    \caption{Schematic of three possible cases of symmetric counterstreams: I. Electrons and ions have
    the same streaming velocity $V_e = V_i$; II. The electron streaming velocity is higher
    than the ion streaming velocity $V_e > V_i$; III. Ions are stationary $V_i = 0$.} \label{f1}%
\end{figure}

We first reanalyze the O-mode instability invoking the recent
results in Refs.~2 and 4. These results are here applied for three
specific cases of symmetric counterstreams with each component
modeled by a drifting bi-Maxwellian distribution function. A
schematic of these plasma systems is presented in Fig. \ref{f1}: I.
The streams are neutral with electrons and ions having the same
streaming velocity $V_e = V_i$; II. The electron streaming velocity
is higher than the ion streaming velocity, $V_e > V_i$; and III.
Ions are at rest, $V_i = 0$, and only electrons are streaming.
Symmetric counterstreams minimize the number of employed parameters,
and enable to analyze the O-mode decoupled from the extraordinary
(X-) mode, which is less susceptible to being unstable. Only
counterstreams of the same species need to be symmetric to satisfy
this condition, but the electron and ion properties (e.g., streaming
velocity, parallel or perpendicular temperature) are not necessary
the same. We are dealing with a single species of ions, namely,
protons.

\subsection{Preliminary conditions: general case}

We assume two-component streams of electron-proton plasmas, and
start their stability analysis from the analytical Eq.~(49) in
Ref.~4
\begin{align} \mu^2 & y^2 + (1+\mu^2)\left[1- {\beta_\parallel \over 2} -P_e
{\mu^2 \over 1+\mu^2} \left(1+{1\over \nu \mu} \right) \right]y
\notag \\ & = \beta_\parallel -1 \left(1 + {\mu \over \nu}
\right)P_e, \label{e0}
\end{align}
with the same streaming parameters
\be \nu \equiv {\epsilon_{e1} \over \epsilon_{p1}}\;
{1-\epsilon_{p1} \over 1-\epsilon_{e1}} \;{V_{e1}^2 \over V_{p1}^2},
\;\;\; P_e \equiv {4 \pi m_e n_0 V_{e1}^2 \over B_0^2} \,
{\epsilon_{e1} \over 1-\epsilon_{e1}}, \nonumber \ee
introduced in Ref.~4, and $y^{1/2} \equiv x_0$ being the limit value
of the squared normalized wavenumber $x_e \equiv k^2 u_{\perp,e}^2 /
(2 \Omega_e^2) = k^2c^2A_e \beta_\parallel / (2 \omega_{pe}^2)$
required by the marginal condition of instability ($\Im(\omega)
\equiv \gamma = 0$), This equation is obtained based on the improved
approximations (36) and (37) in the same reference, and here in the
next will be refined by neglecting $1 \ll \mu^2$ (or $\mu^{-2} \ll
1$), where $\mu = m_p/m_e = 1836$ is the proton-electron mass ratio,
and removing the restriction to very high values of the parameter
$\nu$. Because of the symmetry of the counterstreams of each species
(with the same relative density $\epsilon_{e1}= \epsilon_{e2}=1/2$,
$\epsilon_{p1}=\epsilon_{p2} = 1/2$, and the same streaming velocity
$V_{e1}=V_{e2}=V_e$, $V_{p1}=V_{p2}=V_p$) the quantities $\nu$ and
$P_e$ simplify as follows
\be \nu = {V_{e}^2 \over V_{p}^2}, \;\;\; P_e = {\omega_{pe}^2 \over
\Omega_e^2} \; {V_e^2 \over c^2}. \label{e5} \ee
Also for simplicity, the electron and ion temperatures are assumed
equal ($T_{e,\parallel} = T_{p,\parallel} = T_\parallel$,
$T_{e,\perp} = T_{p,\perp} = T_\perp$), implying
$\beta_{e,\parallel} = \beta_{p,\parallel}=\beta_{\parallel}$.

Equation~(\ref{e0}) then reads
\be y^2 + ay +b = 0, \label{e1} \ee
with
\be a = 1- {\beta_\parallel \over 2} - P_e \left(1+{1 \over \nu \mu}
\right), \ee
\be b = {1 \over \mu^2}\left[1-\beta_\parallel - \left(1+{\mu \over
\nu} \right)P_e \right], \ee
This equation admits a positive solution $y > 0$ when at least one
of the two coefficients $a$ or $b$ is negative. In a low $\beta < 1$
regime this condition is satisfied when the terms depending on $P_e$
are large enough. Thus, $a < 0$ is satisfied if
\be P_e > {1-{\beta_\parallel \over 2} \over 1 + {1 \over \nu \mu}},
\ee
and $b < 0$ if
\be P_e > {1-\beta_\parallel \over 1 + {\mu \over \nu}}.
\label{e7}\ee
For a low $\beta_\parallel < 1$
\be {1-{\beta_\parallel \over 2} \over 1 + {1 \over \nu \mu}} >
{1-\beta_\parallel \over 1 + {\mu \over \nu}}, \ee
leading to the necessary condition $b < 0$, also found in Ref. 4,
Eq. (50). Looking to the solutions of Eq. (\ref{e1}) we can easily
observe that this condition $b < 0$ is also sufficient to have at
least one solution positive
\be y = {-a + \sqrt{a^2 -4b} \over 2}  >0, \label{e9}   \ee
that yields, explicitly,
\begin{align} x_0^2 = & {1 \over 2} \left[P_e \left(1 + {1 \over \nu\mu}\right)
+ {\beta_\parallel \over 2}-1 \right] \notag \\
& + {1 \over 2} \left[ \left({\beta_\parallel \over 2} -1 \right)^2
+P_e^2 \left(1+ {1 \over \nu \mu} \right)^2 - 2 P_e \left(1-
{\beta_\parallel \over 2}\right) \right. \notag \\
& \left.  +{2 P_e \over \nu \mu}\left(1+{\beta_\parallel \over
2}\right)\right]^{0.5}. \label{e10}
\end{align}
Otherwise, for a high $\beta_\parallel > 1$, $b <0 $ is satisfied
for any value of $P_e$, and the same solution (\ref{e9}) remains
positive.

\subsection{Interlude: condition $b < 0$}

Here we analyze in detail the necessary condition $b < 0$ for particular
cases, when only the electrons are counterstreaming and ions are
stationary (forming just a neutralizing background), and for neutral
beams when both the electrons and ions have the same streaming
velocities.

\subsubsection{Counterstreams of electrons}

When ions are at rest, $V_i = 0$, then $\nu \to \infty$, and the
instability condition (\ref{e7}) becomes (also see Eq. (50) from
Ref. 4)
\be P_e > 1- \beta_\parallel \label{e11}. \ee
Using the explicit form in Eq. (\ref{e5}), the instability condition
(\ref{e11}) requires $ V_e^2 > (1-\beta_\parallel) {c^2 \Omega_e^2/
\omega_{p,e}^2} $ or
\be \beta_\parallel > {1 \over 1 + {V_e^2 \over u_{e,\parallel}^2}}
\ee
which is less constrained than the condition derived in Ref.~9
\be \beta_\parallel > {2 \over 1 + {2V_e^2 \over u_{e,\parallel}^2}}
> {1 \over 1 + {V_e^2 \over u_{e,\parallel}^2}} \ee
For cold beams we recover the same condition derived in Ref.~14 $V_e
> c \Omega_e / \omega_{p,e}$.

\subsubsection{Neutral counterstreams}

When both the electrons and ions are streaming with the same
velocity $V_e = V_p = V$, implying $\nu = 1$, the same condition
(\ref{e7}) becomes
\be P_e > {1- \beta_\parallel \over 1 + \mu/\nu} = {1 -
\beta_\parallel \over \mu}. \label{e14} \ee
Writing $P_e$ in terms of Alfven speed $V_A = c \Omega_p /
\omega_{p,p}$
\be P_e \equiv {\omega_{p,e}^2 V_e^2 \over \Omega_e^2 c^2} = {V_e^2
\over \mu V_A^2}\ee
implies in condition (\ref{e14})
\be V > V_A (1 - \beta_\parallel)^{1/2}, \ee
which is the same with condition (7) from Ref.~11. For cold beams we
find necessary $ V> V_A$, the same condition derived in Ref.~15.
Since $V_A = c \Omega_p / \omega_{p,p}= c (m_e /m_p)^{0.5} (\Omega_e
/ \omega_{p,e}) < c \Omega_e / \omega_{p,e}$, it follows that a
system with counterstreams of protons (ions) is much more
susceptible to the instability than one with only counterstreams of
electrons.

\begin{figure*}
\includegraphics[width=80mm]{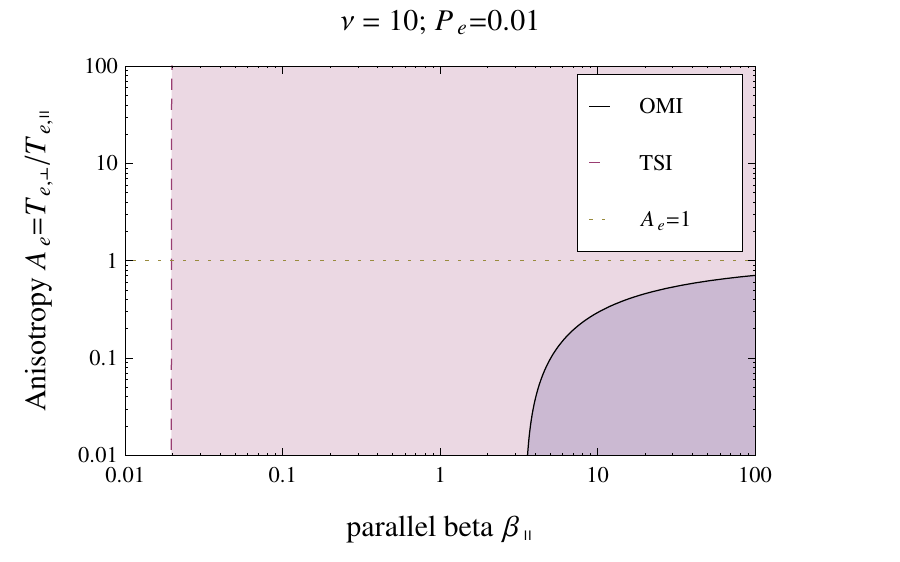}
\includegraphics[width=80mm]{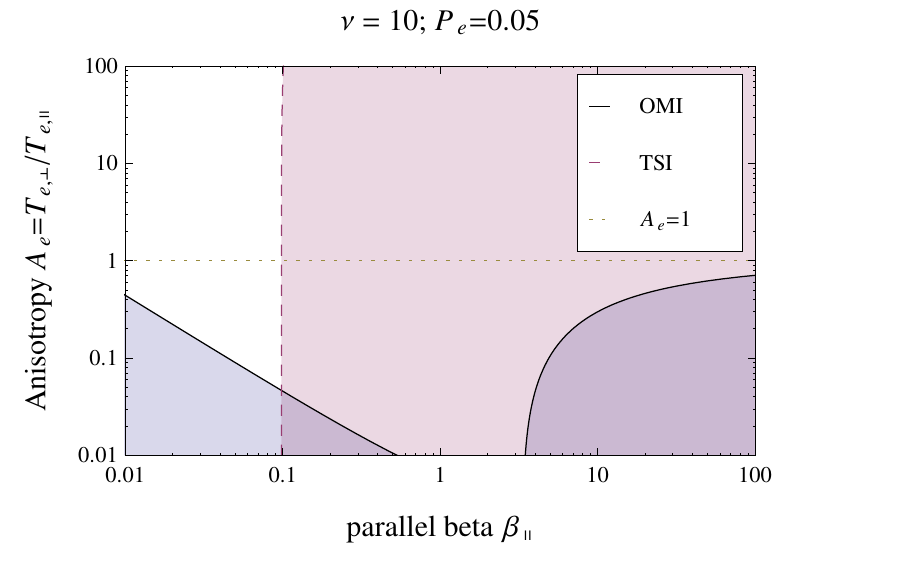}\\
\includegraphics[width=80mm]{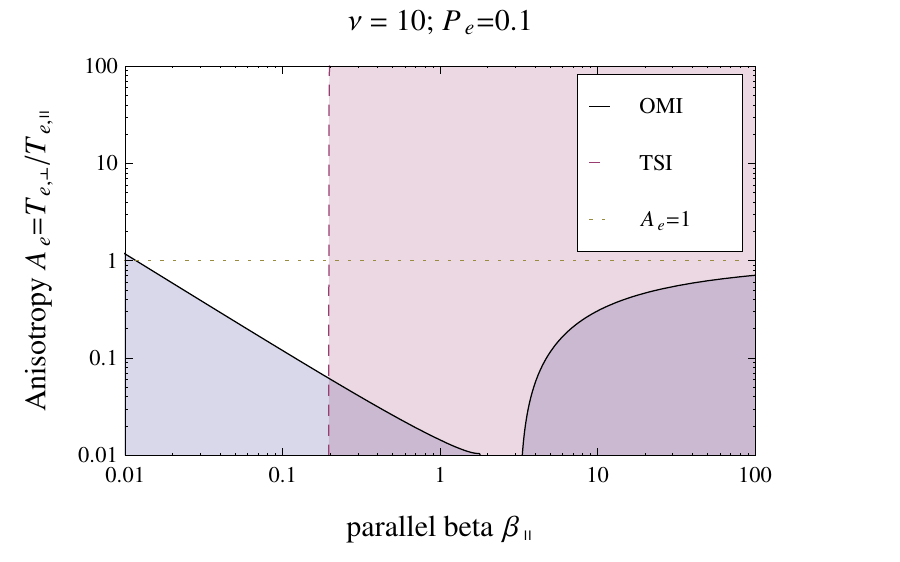}
\includegraphics[width=80mm]{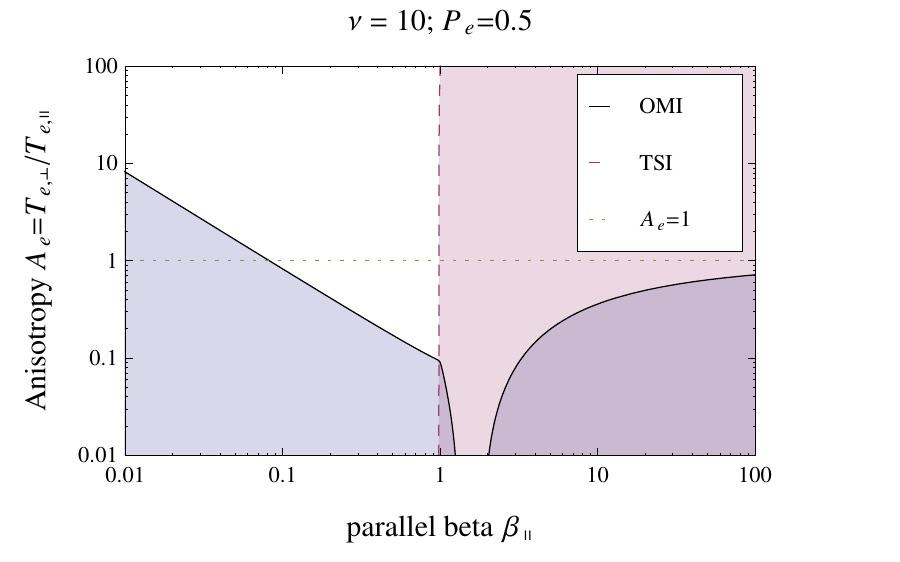}\\
\includegraphics[width=80mm]{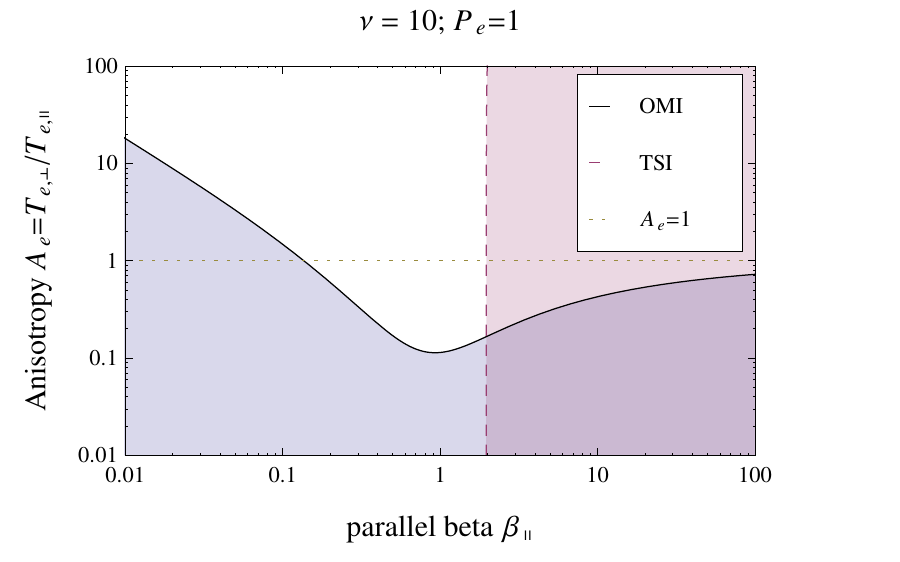}
\includegraphics[width=80mm]{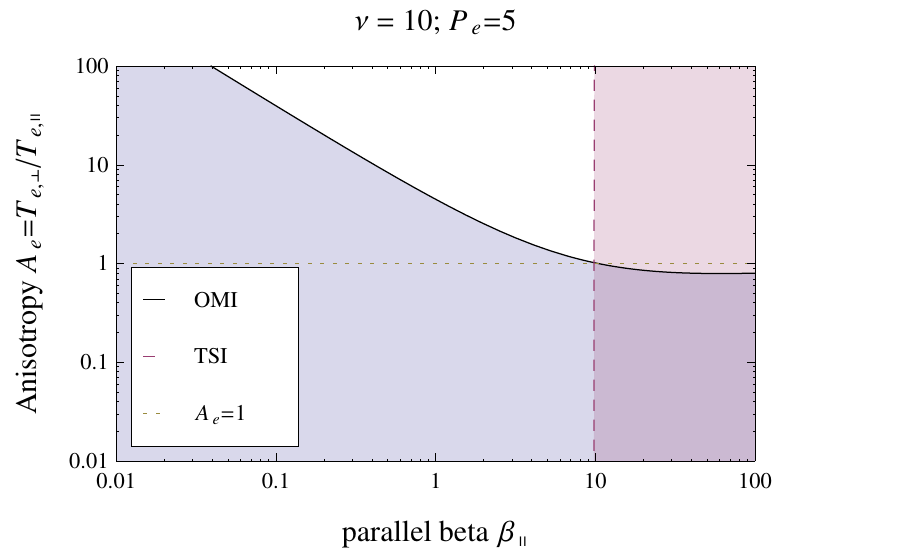}
    \caption{Marginal instability for the OMI from Eq. (\ref{e17}) (solid line), and for the TSI from Eq. (\ref{e26})
    (dashed line), when $\nu = V_e^2 /V_i^2 = 10$. The OMI can develop only in the darkest shading
    where streaming velocity is below the threshold required for the onset of TSI.} \label{f2}%
\end{figure*}

\begin{figure*}
\includegraphics[width=80mm]{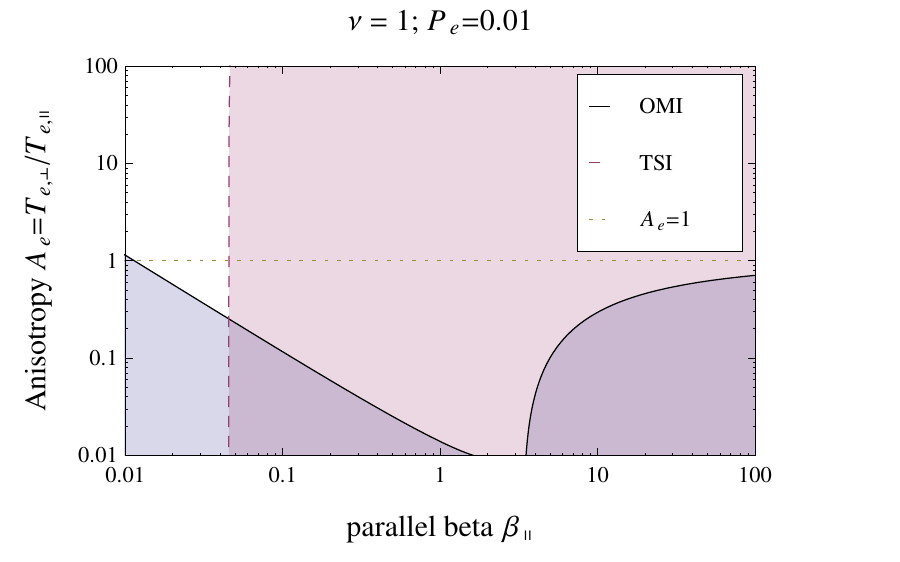}
\includegraphics[width=80mm]{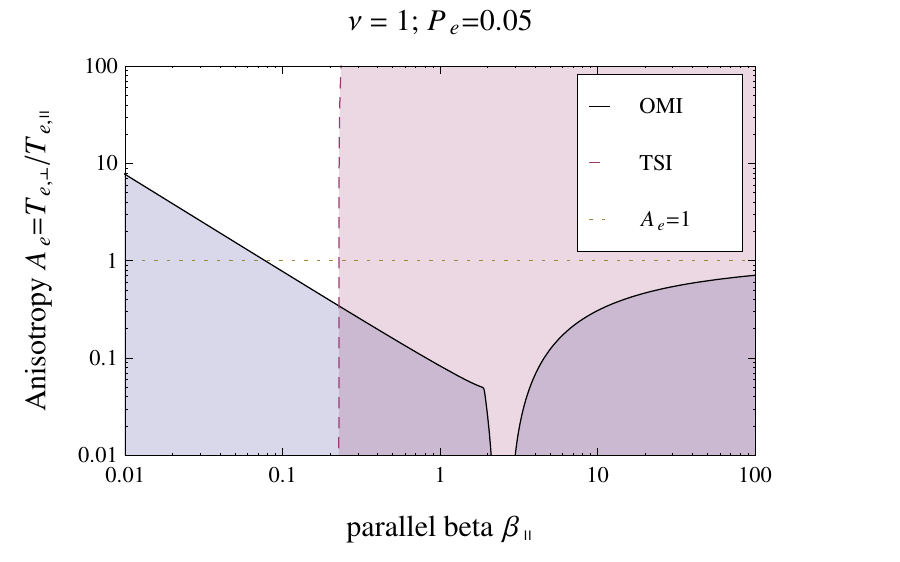}\\
\includegraphics[width=80mm]{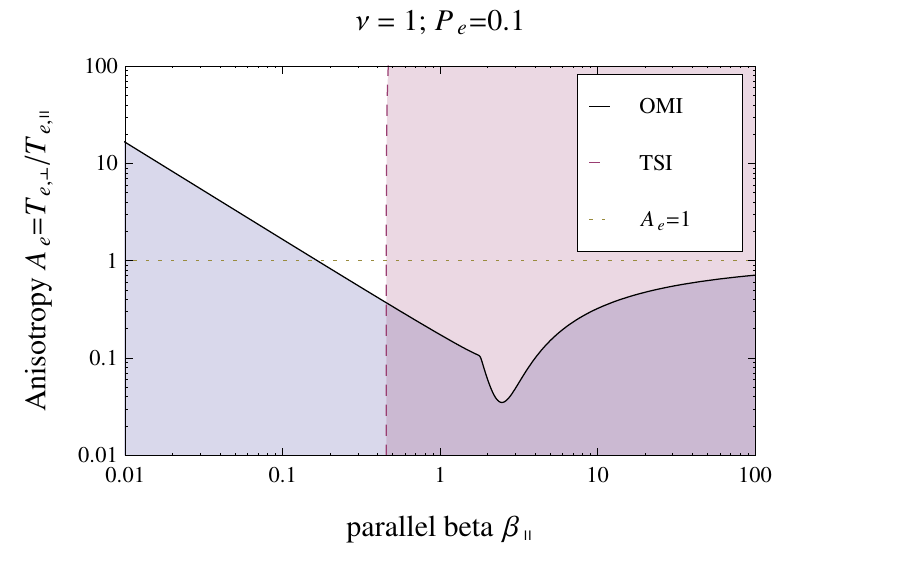}
\includegraphics[width=80mm]{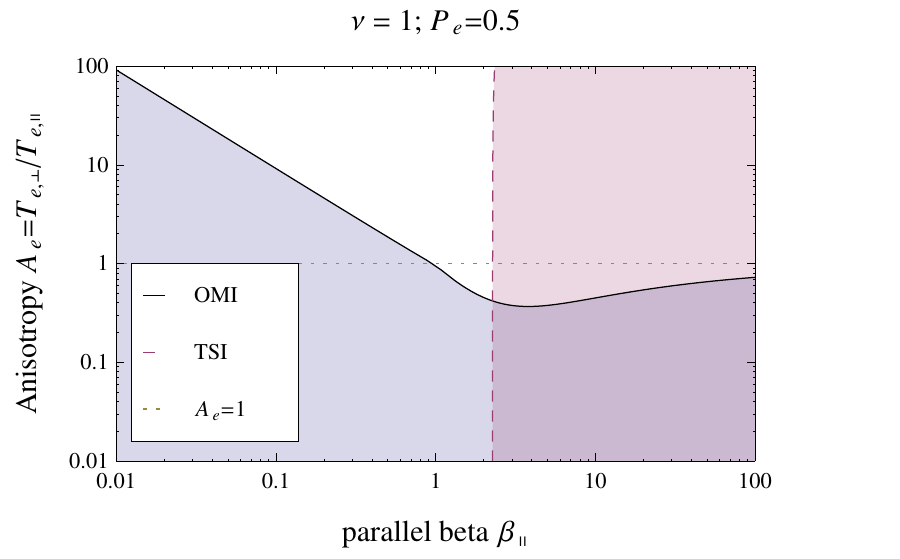}\\
\includegraphics[width=80mm]{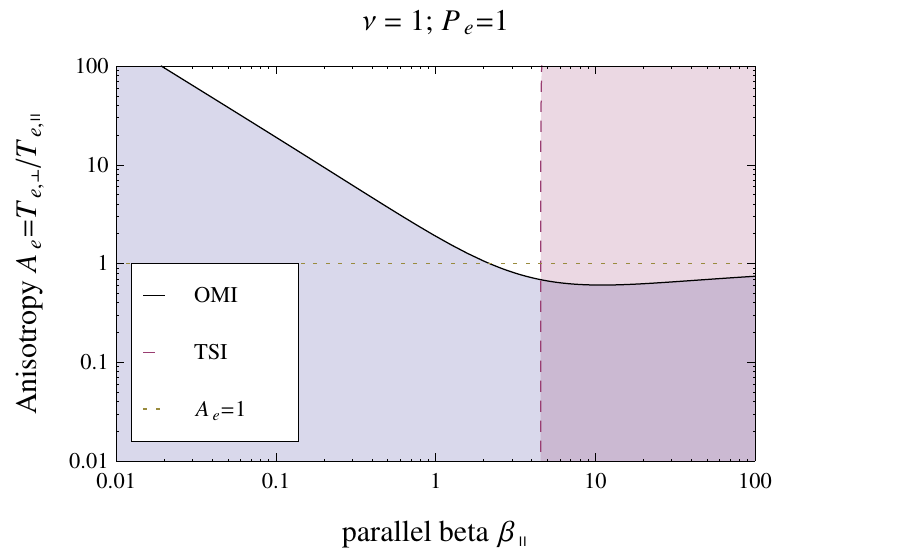}
\includegraphics[width=80mm]{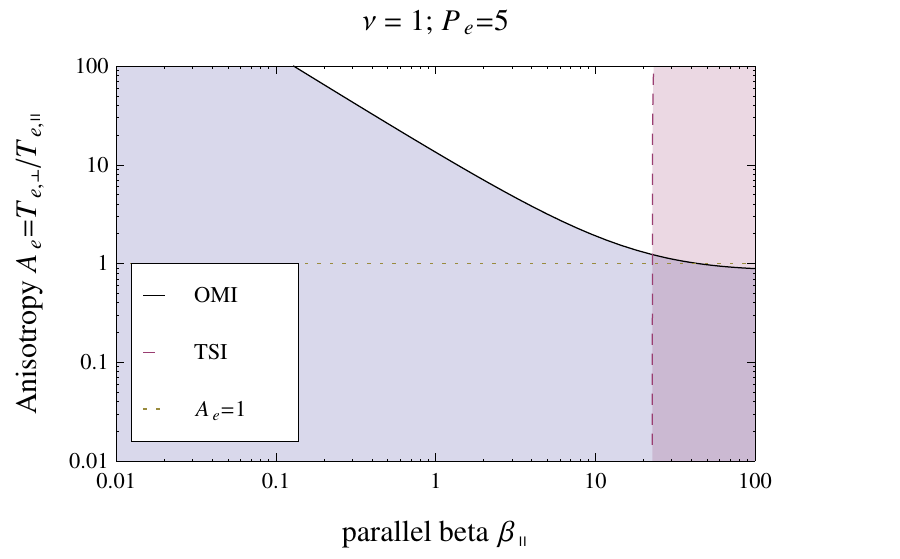}
    \caption{Marginal instability for the OMI from Eq. (\ref{e17}) (solid line), and for the TSI from Eq. (\ref{e26})
    (dashed line) for neutral counterstreams, i.e., $V_e = V_i = V$ ($\nu =1$). The OMI can develop only in the darkest shading
    where streaming velocity is below the threshold required for the onset of TSI.} \label{f3}%
\end{figure*}

\subsection{The instability condition}

With $x_0$ derived in Eq. (\ref{e10}) the marginal instability
condition is readily found from Eq. (34) in Ref.~[4]
\be A < W (x_0) + {W (\mu x_0) \over \mu} + {2P_e \over
\beta_\parallel} \left[W (x_0) + {W (\mu x_0) \over \nu} \right] -
{2x_0 \over \beta_\parallel}, \label{e17} \ee
where $W(z) = 1 - e^{-z}I_0(z)$. We use this condition to derive the
marginal stability against the O-mode instability. This condition is
displayed with solid lines in Figs.~\ref{f2} and \ref{f3} for a
number of relevant cases. The O-mode must be unstable in the gray
shading below the solid lines. In Fig.~\ref{f3} counterstreams are
chosen to be neutral, i.e., with $V_e = V_p$ (case I in
Fig.~\ref{f1}), while in Fig.~\ref{f2} the electron streams are
faster than ions ($V_e > V_p$, i.e., case II in Fig.~\ref{f1}) with
$\nu = 10$.

\begin{figure*}
\includegraphics[width=80mm]{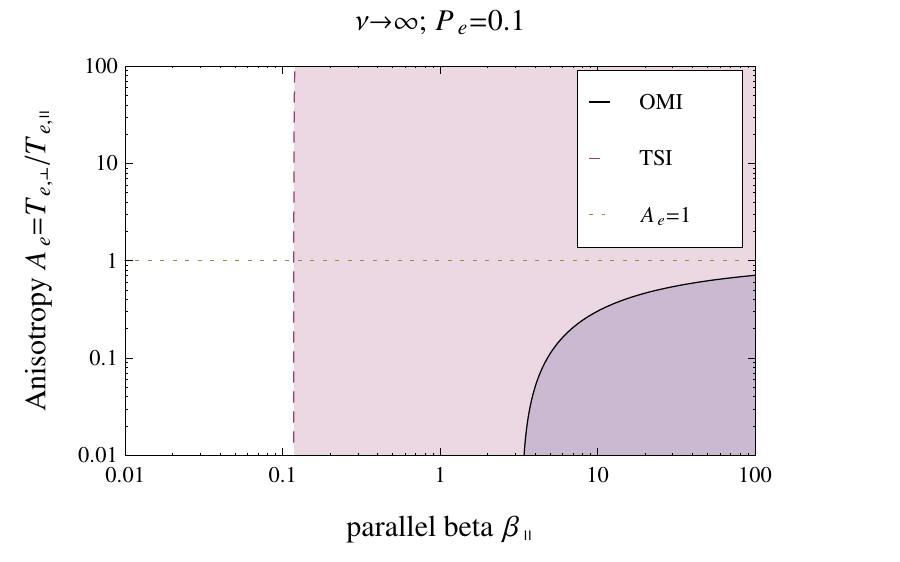}
\includegraphics[width=80mm]{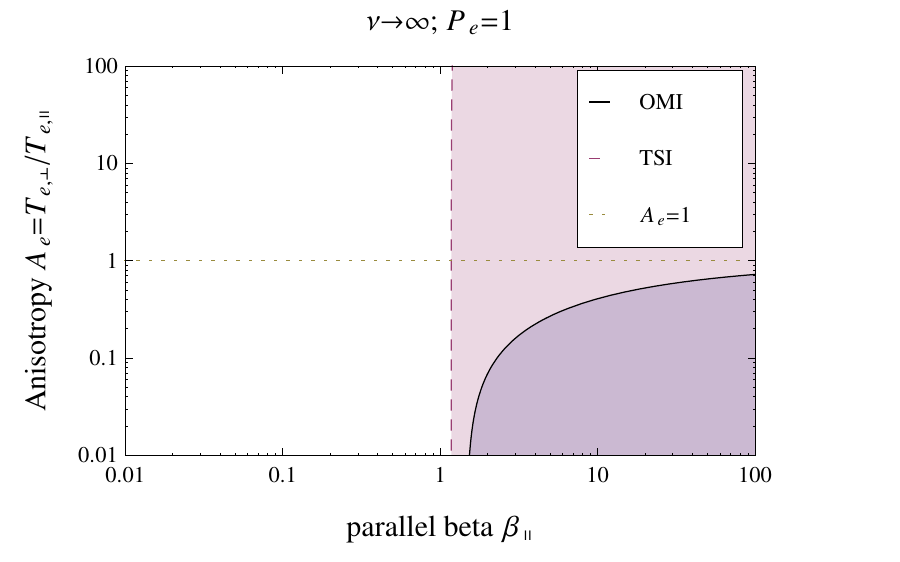}\\
\includegraphics[width=80mm]{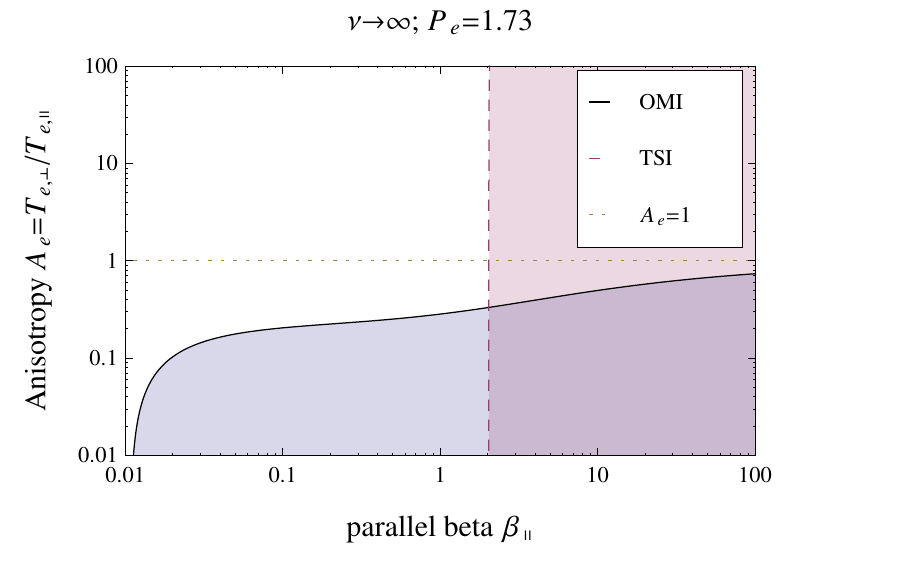}
\includegraphics[width=80mm]{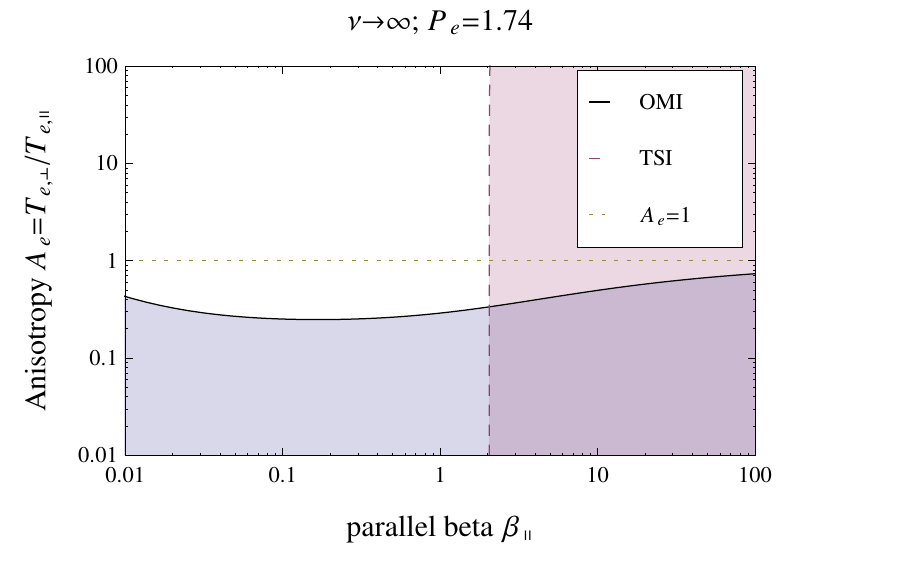}\\
\includegraphics[width=80mm]{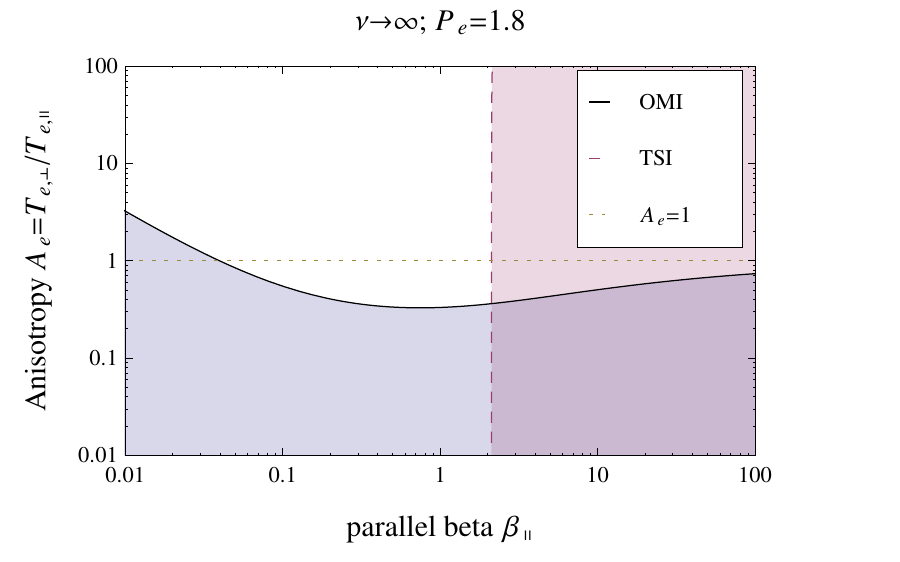}
\includegraphics[width=80mm]{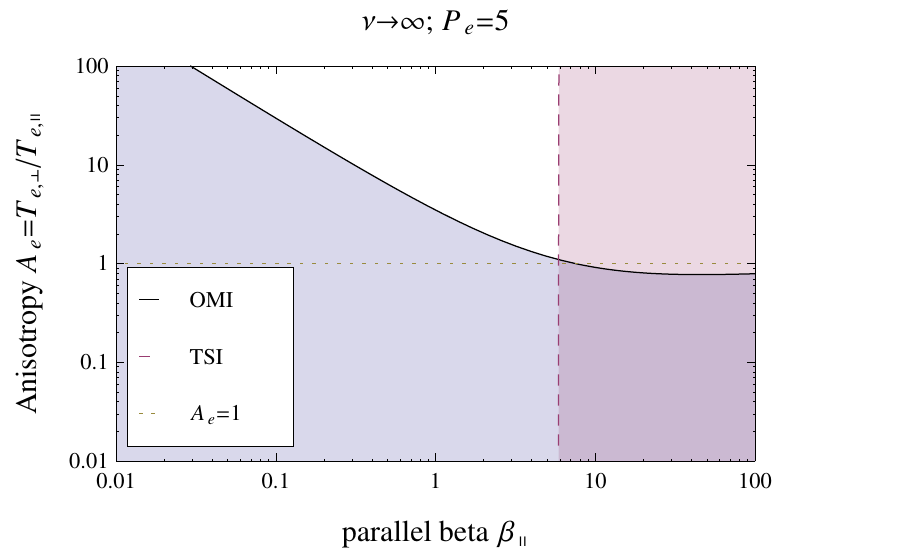}
    \caption{Marginal instability for the OMI, from Eq. (\ref{e19}) (solid line), and for the TSI from Eq. (\ref{e23})
    (dashed line), driven only by the counterstreams of electrons (stationary
    ions, $V_i = 0$). The OMI can develop only in the darkest
    shading where the streaming velocity is below the threshold required for the onset of TSI.} \label{f4}%
\end{figure*}

For the third case, when ions are stationary and only electrons are
counterstreaming (case III in Fig. \ref{f1}), the parameter $\nu \to
\infty$ becomes very large, reducing the expression in
Eq.~(\ref{e10})
\begin{align} x_0^2 = P_e + {\beta_\parallel \over 2}-1, \label{e18}
\end{align}
and the instability condition from Eq.~(\ref{e17})
\be A < W (x_0) + {W (\mu x_0) \over \mu} + {2 \over
\beta_\parallel} [P_e W (x_0) -x_0]. \label{e19} \ee
Notice that by comparison to Ref.~2, where the instability condition
is derived by neglecting the ion effects, here the ion nonstreaming
effects are still present in the right-side (second term) of
inequality (\ref{e19}). This new form in Eq.~(\ref{e19}) is used in
the present paper to derive the marginal stability of the electron
counterstreams, which is displayed with solid lines in
Fig.~\ref{f4}.

\section{Interplay with the electrostatic two-stream instability}

Counterstreaming plasmas are also subject to the electrostatic
streaming instabilities, i.e., electron-electron ($e-e$),
electron-proton ($e-p$), and proton-proton ($p-p$) instabilities,
with a maximum growth in the streaming direction. Of these, the most
efficient (or faster) are the instabilities driven by electrons,
i.e., $e-e$ or $e-p$ two-stream instabilities \cite{st64,la02}.
Notice that no acoustic mode can be excited if the plasma
populations and components are isothermal ($T_e \sim T_p$). The
two-stream instability is also faster than the O-mode instability
\cite{le71, ga72}, except for streaming velocity very near or below
the threshold for the onset of the two-stream instability. The
electrostatic instability is only inhibited by the thermal spread of
plasma particles in the streaming direction (i.e., the parallel
temperature), and it is therefore expected to become even more
competitive against the O-mode instability in the low $\beta < 1$
regime, where either the two-stream instability is enhanced by the
low temperature, or the O-mode instability is inhibited by the
intense magnetic field \cite{st06,bo70,le71,ga72}.

For that reason, the existence of the O-mode instability can be well
established only below the marginal condition for the two-stream
instability. Here we propose to delimitate these regimes based on
the results in Ref. [13], where the marginal condition has been
derived systematically for different two-stream instabilities. The
symmetry conditions imposed in our present study for the
counterstreaming plasmas (enables decoupling of the O-mode from the
X-mode) along with $T_e \simeq T_p$ (ubiquitous in space plasmas)
lead to a reduced number of two-stream instabilities specified in
Table \ref{t1}. Relevant for us here are only the plasma
counterstreams $a - b$ of the same species $a=b=e,p$ or different
species $a=e$, $b=p$, but satisfying
\be \Lambda \equiv \left(\epsilon_a T_b \over \epsilon_b T_a
\right)^{1/2} =1, \label{e20} \ee
(since $\epsilon_a = \epsilon_b = 1/2$, and $T_a = T_b$, see
previous section). In this case the marginal condition of
\emph{stability} for two generic counterstreams $a - b$ is
\cite{la02}
\be {|\vec{V}_b - \vec{V}_a| \over u_{b,\parallel}} \leqslant 0.92
\left(1 + {\omega_{p,b}\over \omega_{p,a}}\right), \label{e21} \ee
We analyze this condition for each type of two-stream instability
found relevant in Table \ref{t1}, and then compare them to derive
the lowest threshold condition.

\subsection{Electron-electron streaming instability}

In this case $a = b =e$, the relative velocity reads $|\vec{V}_b -
\vec{V}_a| = 2 V_e$ and the marginal condition (\ref{e21}) becomes
\cite{st64}
\be V_e \leqslant 0.92 \; u_{e,\parallel}. \label{e22} \ee
Now, to adapt within a $A\varpropto \beta_\parallel^{-1}$-dependence
(temperature anisotropy vs. inverse plasma beta) we use definition
(\ref{e5}), and express condition (\ref{e22}) in terms of the
(electron) streaming parameter $P_e$ and the parallel plasma beta
\be P_e \leqslant 0.85 \; \beta_{\parallel}. \label{e23} \ee

\subsection{Proton-proton streaming instability}

If the instability is driven only by protons, i.e., $a = b =p$, the
marginal condition (\ref{e21}) provides a lower threshold for the
streaming velocity of protons (also discussed by Stringer
\cite{st64})
\be V_p \leqslant 0.92 \; u_{p,\parallel}. \label{e24} \ee
since $u_{p,\parallel} = u_{e,\parallel} / \mu^{1/2} <
u_{e,\parallel}$ ($\mu = m_p/m_e = 1836$).
The first two cases in Fig. \ref{f1} include conditions when the
proton streaming velocity is high enough to drive the two-stream
$p-p$ instability. However, developing of this instability (with a
growth rate of the order of proton plasma frequency) is not
realistic since the O-mode instability (with a growth rate of the
order of electron gyrofrequency) is expected to be much faster.

\subsection{Electron-proton streaming instability}

In this case a Buneman-like instability is induced
\cite{bu59,st64,la01}, and the lowest threshold is found from the
same condition (\ref{e21}) if we consider the case $a = p$ and $b
=e$
\be V_e \leqslant 0.92 \; u_{e,\parallel} \; {1 + {1\over \mu^{1/2}}
\over 1 \pm {1\over \nu^{1/2}}}. \label{e25} \ee
Furthermore, we can distinguish between two growing modes
distinctively assigned to "$\pm$" in the denominator. Thus, the
instability can be driven by electrons and protons streaming either
in the same direction, when the relative drifting speed is only
$|\vec{V}_e - \vec{V}_p| = V_e - V_p$, or in opposite directions,
when the relative drifting speed is higher $|\vec{V}_e - \vec{V}_p|
= V_e + V_p$ and determines a lower threshold, asigned to "$+$" in
the denominator in Eq. (\ref{e25}). In terms of the streaming
parameter $P_e$ and the parallel plasma beta this condition reads
\be P_e \leqslant 0.85 \; \beta_{\parallel} \; \left({1 + {1\over
\mu^{1/2}} \over 1 + {1\over \nu^{1/2}}}\right)^2. \label{e26} \ee

Now, comparing conditions (\ref{e23}) and (\ref{e26}), we find that
the lowest threshold is defined by (\ref{e23}) only if $\nu > \mu =
1836$, condition well satisfied when ions are almost stationary,
e.g., the limit case III. Indeed, in the limit case III, the ions
are stationary $\nu \to \infty$, and the two-stream $e - e$
instability presents the lowest marginal condition given by
(\ref{e23}). In the other limit of neutral counterstreams, when the
electrons and ions move with the same streaming velocity, $\nu = 1$
and the lowest marginal condition is that against the
electron-proton instability reduced to
\be P_e \leqslant 0.21  \; \beta_{\parallel} \; \left(1 + {1\over
\mu^{1/2}}\right)^2 \simeq 0.22 \; \beta_{\parallel} < 0.92 \;
\beta_{\parallel}. \label{e27} \ee
The lowest marginal conditions of stability against the
electrostatic two-stream instabilities are summarized in Table
\ref{t1}, indicating three different cases, which are not
necessarily related to our first classification, i.e., cases I, II
and III.

\begin{table}
\caption{\label{t1} Types of electrostatic two-stream instabilities
relevant for our cases I, II, and III, and the lowest marginal
condition of stability.}
\begin{ruledtabular}
\begin{tabular}{llll}
& $\nu$ & Instabilities & Marginal condition \\
\hline I. & $\nu =1$ & $e-e$, $e-p$, $p-p$ &  Eq. (\ref{e27}) \\
II. & $\nu < \mu $ & $e-e$, $e-p$, $p-p$ & Eq. (\ref{e26}) \\
 & $\nu > \mu $ & $e-e$, $e-p$, $p-p$ & Eq. (\ref{e23}) \\
III. & $\nu \to \infty$ & $e-e$, $e-p$ & Eq. (\ref{e23})\\
\end{tabular}
\end{ruledtabular}
\end{table}

Thus, a new distinction can be made function of the parameter $\nu$
as it takes values less or higher than the proton-electron mass
ratio $\mu$. If $\nu < \mu$ is satisfied the lowest marginal
condition against any electrostatic instability is given by
(\ref{e26}). For an arbitrary value $\nu = 10$ (case also studied in
Ref.[4]) this condition becomes $P_e \leqslant 0.51\;
\beta_\parallel$, and it is displayed with dashed lines in
Fig.~\ref{f2}. For the limit case when $\nu = 1$, the same marginal
condition simplifies to (\ref{e27}), which is displayed with dashed
lines in Fig.~\ref{f3}. In the opposite case, when $\nu > \mu$ is
satisfied, including the limit case $\nu \to \infty$ of stationary
ions, the lowest marginal condition against any electrostatic
instability is given by (\ref{e23}). This condition is displayed
with dashed lines in Fig.~\ref{f4}.

The regimes where only the O-mode instability can operate are always
found in the right-hand side of these dashed lines, i.e., the
lighter gray shading. The superposition with the conditions for
O-mode instability indicates for the existence of this instability
only the darkest shading regions. In the high beta ($\beta_\parallel
>1$) plasmas the existence of the O-mode instability is not
significantly affected, unless the streams are very energetic ($P_e
> 1$). But the existence of this instability is drastically
restrained by the interplay with the two-stream instability in the
low beta ($\beta_\parallel <1$) regimes.

\section{Discussion and conclusions}

We have studied the marginal conditions for the O-mode instability
by contrast to those of the electrostatic instabilities. In the
process of relaxation of the counterstreaming plasmas the
electrostatic two-stream instabilities (with a growth rate of the
order of electron plasma frequency) are usually much faster than the
O-mode instability (with a growth rate of the order of electron
gyrofrequency). The existence of the O-mode can therefore be
established only for streaming velocities below the threshold of the
two-stream instabilities.

Our present analysis is based on accurate analytical expressions of
the marginal conditions of instability provided in Ref. 4 for the
O-mode instability, and in Ref. 13 for the two-stream instabilities.
The refined analysis in Secs. II and III aims to assess the
robustness of these analytical expressions by the agreements found
with particular cases studied before, e.g., stationary ions, neutral
streams, $e-e$ or $e-p$ counterstreams. Despite the limitations
imposed by the symmetry of the counterstreams, this seems to be the
most convenient way to study the O-mode (decoupled from other plasma
modes) and make its properties more transparent. Presently there is
an increased interest for the O-mode instability, especially for
understanding its activity in the low-beta plasmas ($\beta < 1$) and
for temperature anisotropies $A = T_\perp / T_\parallel$ even larger
than unity. As a mechanism of limitation of the kinetic
anisotropies, this instability could provide a plausible explanation
for the low-beta boundaries of stable plasma configurations observed
in the solar wind and terrestrial magnetosphere.

Our investigation on the competition with the two-stream
instabilities reveals that the parameter range of the O-mode
instability is significantly restrained, especially in the low-beta
plasmas. Thus the low-beta regimes where only the O-mode instability
can operate are significantly restrained by a minimum cutoff, given
by
\be \beta_{\parallel} > \beta_{\parallel,c} \equiv {P_e \over 0.85}
\; \left({1 + {1 \over \nu^{1/2}} \over 1 +{1 \over
\mu^{1/2}}}\right)^2 \label{e28} \ee
if $\nu < \mu = 1836$ is satisfied, see Figs.~\ref{f2} and \ref{f3},
or by
\be  \beta_{\parallel} > \beta_{\parallel,c} \equiv {P_e \over 0.85}
 \label{e29} \ee
in all the other cases, see Fig.~\ref{f4}. The activity of this
instability in the high-beta plasmas can be also constrained by the
two-stream instability, if $P_e$ is large enough, see bottom panels
in Figs.~\ref{f2} and \ref{f3}.

Moreover, the existence of the O-mode instability becomes limited
only to sufficiently small $A = T_\perp /T_\parallel$, less than a
maximum value $A_{m}$ given by Eqs. (\ref{e17}) and (\ref{e10}) at
the plasma beta cutoff ($\beta_{\parallel,c}$) derived above, i.e.,
\be A_{m} = W(x_c) + {W(\mu x_c) \over \mu} + {2P_e \over
\beta_{\parallel,c}} \left[W(x_c) + {W(\mu x_c) \over \nu} \right] -
{2x_c \over \beta_{\parallel,c}} \label{e30} \ee
with
\begin{align} x_c^2 = & {1 \over 2} \left[P_e \left(1 + {1 \over \nu\mu}\right)
+ {\beta_{\parallel,c} \over 2}-1 \right] \notag \\
& + {1 \over 2} \left[ \left({\beta_{\parallel,c} \over 2} -1
\right)^2 +P_e^2 \left(1+ {1 \over \nu \mu} \right)^2 \right. \notag \\
& \left. - 2 P_e \left(1-{\beta_{\parallel,c} \over 2}\right) +{2
P_e \over \nu \mu}\left(1+{\beta_{\parallel,c} \over
2}\right)\right]^{0.5}. \label{e31}
\end{align}
Estimations can be made if we, for instance, take a few examples of
a less particular case when $\nu < \mu$, like the ones displayed in
Fig. \ref{f2} ($\nu = 10$). In this case $\beta_{\parallel, c}
\simeq 1.96 P_e$ and $x_c^2 = 0.5(3.92 P_e^2 - 3.96 P_e +1)^{0.5} +
0.99 P_e -0.5 \simeq 2P_e-1$, and the limit values $A_m$ calculated
with Eq. (\ref{e30}) are listed in Table \ref{t2} for different
values of the parameter $P_e$. Notice that values larger than unity
$A_m >1$ for the temperature anisotropy are admitted, but only for
sufficiently large electron streaming velocities corresponding to
$P_e > 1$, and only for sufficiently large values of the plasma beta
parameter $\beta_{\parallel} > 1$.

\begin{table}
\caption{\label{t2} Limit (maximum) values of the temperature
anisotropy ($A_m$) given by Eq.~(\ref{e30}) for the cases displayed
in Fig.~\ref{f2}.}
\begin{ruledtabular}
\begin{tabular}{llllll}
$P_e$ & 0.05  & 0.1 & 0.5 & 1.0 & 5.0\\
\hline $\beta_{\parallel,c}$ & 0.098 & 0.19 &  0.98 & 1.96 & 9.8 \\
$x_c (\times 10^{-2})$ & 0.17 & 0.26 &  4.72 & 99.0 & 298 \\
$A_m$ & 0.047 & 0.062 & 0.094 & 0.170 & 1.02 \\
\end{tabular}
\end{ruledtabular}
\end{table}

To conclude, the conditions where the O-mode instability can operate
efficiently are markedly constrained by the electrostatic
instabilities in both the low $\beta_\parallel < 1$ and large
$\beta_\parallel > 1$ plasmas. Thus the existence of the O-mode
instability becomes possible only for sufficiently high values of
the plasma beta parameter in parallel direction $\beta_{\parallel}
\geqslant \beta_{\parallel,c}$. This restriction is particularly
important for the low-beta plasmas, since the existence of this
instability was claimed in the previous studies for any small value
(without limit) of the parallel plasma beta $\beta_{\parallel} \to
0$. The instability of the O-mode can develop in the low-beta
plasmas but only for $\nu < \mu$ and a streaming energy density less
than the magnetic energy density ($P_e < 1$). Further comparison is
now possible with the solar wind streaming conditions to establish
whether this instability can explain the observed limits of the
anisotropy or not. For a supra-unitary anisotropy $A = T_\perp
/T_\parallel > 1$, the stimulation of this instability by the
energetic streams seems to be impossible (or unrealistic) in the
low-beta plasmas. These situations appear to be resolved by a more
realistic dissipation of the streaming free energy by the
electrostatic instabilities.


\begin{acknowledgements}
The authors acknowledge support from the Katholieke Universiteit
Leuven, Grant nr. SF/12/003, and from the Ruhr-Universit\"at Bochum,
the Deutsche Forschungsgemeinschaft (DFG), Grants Schl 201/25-1 and
SH 21/3-2. These results were obtained in the framework of the
projects GOA/2015-014 (KU Leuven), G.0729.11 (FWO-Vlaanderen) and
C~90347 (ESA Prodex 9). The research leading to these results has
also received funding from the European Commission's Seventh
Framework Programme (FP7/2007-2013) under the grant agreements
SOLSPANET (project n° 269299, www.solspanet.eu) and eHeroes
(project n° 284461, www.eheroes.eu).
\end{acknowledgements}

\end{document}